\documentclass[aps,twocolumn,groupedaddress]{revtex4}
\usepackage{epsfig}
\usepackage{graphicx}
\usepackage[T1]{fontenc}
\usepackage{ae}
\usepackage{color}
\usepackage[latin1]{inputenc}
\usepackage{amssymb,amsbsy,amsmath}
\usepackage{bbm}
\usepackage{undertilde}

\newtheorem{defi}{Definition}
\newtheorem{prop}{Proposition}
\newtheorem{ass}{Assumption}

\begin{document}



\title[Sequential measurements and entropy]
{Sequential measurements and entropy}

\author{Heinz-J\"urgen Schmidt and Jochen Gemmer
}
\address{ Universit\"at Osnabr\"uck,
Fachbereich Physik,
 D - 49069 Osnabr\"uck, Germany
}


\begin{abstract}
We sketch applications of the so-called J-equation to quantum information theory concerning
fundamental properties of the von Neumann entropy. The J-equation has recently be proposed
as a sort of progenitor of the various versions of the Jarzynski equation. It has been derived
within a general framework of sequential measurements that is slightly generalised here.
\end{abstract}

\maketitle

\section{Introduction}\label{sec:I}

Entropy and its increase in closed systems, the so-called $2^{nd}$ law, historically arose in classical thermodynamics and statistical physics
of many-body systems.
The status of these concepts in quantum theory is not completely clear despite the vast amount of literature on this subject.
The time-honoured definition of the von Neumann entropy $S(\rho)=-\mbox{Tr}(\rho\,\log\rho)$ has the following properties:
It remains constant under unitary time evolution and increases (always understood in the sense of including the case of
remaining constant) during projective measurement \cite{vN32}:
\begin{equation}\label{I1}
 S(\rho)\le S\left(\sum_n \utilde{P}_n\,\rho\,\utilde{P}_n\right)
 \;,
\end{equation}
with self-explaining notation. These properties suggest to relate the $2^{nd}$ law and, more generally,
the basic concepts of quantum thermodynamics to sequential measurements, real or hypothetical ones.

A successful approach following these lines of thought has lead to the  Jarzynski equation
\begin{equation}\label{I2}
 \left\langle e^{-\beta \,w}\right\rangle = e^{-\beta \Delta F}
 \;,
\end{equation}
that is the most famous representative of a class of similar fluctuation theorems \cite{J97} -- \cite{TMYH13}.
The Jarzynski equation is an exact statement about the expectation value of a non-linear function of
the work $w$, viewed as the energy difference of two sequential energy measurements.
Between the two measurements an arbitrary unitary time evolution takes place.
Although ``work" is not an observable in the traditional sense of a self-adjoint operator
\cite{TLH07}
it can be understood as an example of the generalised observable concept described by a
positive operator valued measure, see \cite{BLPY16} -- \cite{CRP15}.
We have performed an analysis \cite{SG20} of (\ref{I2}) and more general variants of the Jarzynski equation
with the result that it can be derived from an equation, called J-equation, that concerns the statistics of
sequential measurements and is initially independent of any realization in quantum theory. From the J-equation
one derives in the usual way, via Jensen's inequality, an inequality that resembles the $2^{nd}$ law.
However, some caution is advised:  The J-equation contains an undetermined probability distribution
that can be chosen in such a way as leading to the Jarzynski equations or, alternatively, to an approach
initially considered by W. Pauli \cite{P28} in the context of time-dependent perturbation theory (``Golden Rule").
Only the second choice gives a proper account of the $2^{nd}$ law. A more general version of this approach
has been published three years later by O.~Klein \cite{K31} and is since known as ``Klein's inequality", see \cite{NC00},
although detached from the thermodynamic context. We will see in Section \ref{sec:GT}
that it can also be derived from the J-equation.

Thus we come across the finding that the $2^{nd}$ law has some aspects that can be viewed as statements
about sequential measurements and are independent of many-body physics. One could object to this viewpoint
that measurements are only possible by interactions with a macroscopic measuring device which
in turn brings many-particle aspects into play.
Without conclusively clarifying these issues, we note that applications of the J-equation result
in a domain that could be seen as a pre-theory of quantum information theory and is, e.~g.,
covered by chapter $11$ of \cite{NC00}. These applications are the subject of the present work.\\

However, we have to slightly extend the mathematical framework presented in \cite{SG20},
in order to include, for example, also statistical operators with the eigenvalue $0$. This
is done in Section \ref{sec:SM}. The next Section \ref{sec:G} deals with realisations of
the statistical model of sequential measurements in quantum theory,
first in general, see Section \ref{sec:GS}, and then
in a special form tailored for current purposes, see Section \ref{sec:GT}.
In addition to the mentioned Klein's inequality we also prove the statement (\ref{I1}) within our framework.
One may ask: What is the purpose of proving familiar propositions anew? The obvious rationale is to
uncovering unexpected relationships between seemingly disjoint domains as non-equilibrium quantum statistics
and quantum information theory. These relationships could, hopefully, also be used to obtain new results
or simplified proofs of known ones, which is, however, beyond the scope of the present article.
In the summary and outlook Section \ref{sec:SO} we will shortly hint at these possibilities.

\section{Statistical model of sequential measurements}\label{sec:SM}

We consider two sequential measurements at the same physical system at times $t_0<t_1$ with respective outcome sets ${\mathcal I}$ and ${\mathcal J}$.
These sets are assumed to be finite or countably infinite.
Hence the joint outcome of the two measurements can be represented
by the pair $(i,j)\in {\mathcal I}\times {\mathcal J}$. We define
\begin{equation}\label{SM1}
 {\sf E}\equiv {\mathcal I}\times {\mathcal J}
\end{equation}
as the set of ``elementary events". The probability of elementary events will be obtained by means of some auxiliary functions
$\Pi,x,\widetilde{x}$
that have no direct statistical meaning. We assume the existence of the functions
\begin{eqnarray}
\label{SM2a}
  \Pi &:& {\sf E}\rightarrow {\mathbbm R}_\ge\;, \\
  \label{SM2b}
  x&:&{\mathcal I}\rightarrow {\mathbbm R}_\ge\;, \\
  \label{SM2c}
   \widetilde{x}&:&{\mathcal J}\rightarrow {\mathbbm R}_\ge\;,
\end{eqnarray}
where ${\mathbbm R}_\ge$ denotes the set of non-negative reals. $\Pi$ will be called the ``conditional matrix"
and its entries are written as $\Pi(j|i)$. Further, the $x(i)$ and  $\widetilde{x}(j)$ will be called
``abstract eigenvalues" of the first and second kind for reasons that will become clear in the next Section.

The marginal sums of $\Pi$ will be denoted by
\begin{eqnarray}
\label{SM3a}
  d(i) &\equiv& \sum_{j\in{\mathcal J}}\Pi(j|i)\le \infty\;,\quad i\in{\mathcal I}\;, \\
  \label{SM3b}
  \widetilde{d}(j) &\equiv& \sum_{i\in{\mathcal I}}\Pi(j|i)\le \infty\;,\quad j\in{\mathcal J}\;,
\end{eqnarray}
and may assume values in ${\mathbbm R}_\ge \cup \{\infty\}$.

Further, we define for all $(i,j)\in{\sf E}$
\begin{eqnarray}
\label{SM4a}
 P(i,j) &\equiv& \Pi(j|i)\,x(i)\;, \\
 \label{SM4b}
 \widetilde{P}(j,i) &\equiv& \Pi(j|i)\,\widetilde{x}(j)\;,
\end{eqnarray}
and postulate our central axiom as
\begin{ass}\label{axiom}
 \begin{equation}
 \label{SM5a}
   \sum_{(i,j)\in{\sf E}}P(i,j) = 1\;,
 \end{equation}
  and
  \begin{equation}
 \label{SM5b}
   \sum_{(i,j)\in{\sf E}}\widetilde{P}(j,i) =1\;.
 \end{equation}
\end{ass}
Eq.~(\ref{SM5a}) especially means that $x(i)=0$ if $d(i)=\infty$, analogously Eq.~(\ref{SM5b}) has to be understood as
$\widetilde{x}(j)=0$ if $\widetilde{d}(j)=\infty$. Both functions $P(i,j)$ and $\widetilde{P}(i,j)$ can be used
to describe probabilities of elementary events. Correspondingly, we obtain the following four marginal probabilities
\begin{eqnarray}
 \label{SM6a}
 p(i) &\equiv& \sum_{j\in{\mathcal J}}P(i,j)\stackrel{(\ref{SM4a})}{=} \sum_{j\in{\mathcal J}}\Pi(j|i)\,x(i)\stackrel{(\ref{SM3a})}{=}d(i)\,x(i)\;,\\
  \label{SM6b}
 q(j) &\equiv& \sum_{i\in{\mathcal I}}P(i,j)\stackrel{(\ref{SM4a})}{=}  \sum_{i\in{\mathcal I}}\Pi(j|i)\,x(i)\;,\\
  \label{SM6c}
 \widetilde{p}(j) &\equiv& \sum_{i\in{\mathcal I}} \widetilde{P}(j,i)\stackrel{(\ref{SM4b})}{=}  \sum_{i\in{\mathcal I}}\Pi(j|i)\, \widetilde{x}(j)\stackrel{(\ref{SM3b})}{=}  \widetilde{d}(j)\, \widetilde{x}(j)\;,\\
 \label{SM6d}
 \widetilde{q}(i) &\equiv& \sum_{j\in{\mathcal J}}\widetilde{P}(j,i)\stackrel{(\ref{SM3b})}{=} \sum_{j\in{\mathcal J}}\Pi(j|i)\,\widetilde{x}(i)\;,
\end{eqnarray}
where $p(i)=d(i)\,x(i)$ has to be set to $0$ if $d(i)=\infty$, analogously for $\widetilde{p}(j)=\widetilde{d}(j)\, \widetilde{x}(j)$.
According to Assumption \ref{axiom} all four marginal probabilities sum to unity.

It may be instructive to calculate the conditional probability belonging to $P(i,j)$, where
we preliminary restrict ourselves to the case $p(i)>0$ for all $i\in{\mathcal I}$:
\begin{equation}\label{SM7}
\pi(j|i)\equiv \frac{1}{p(i)}P(i,j)\stackrel{(\ref{SM4a})}{=}\frac{1}{p(i)}\Pi(j|i)\,x(i)\stackrel{(\ref{SM6a})}{=}\frac{1}{d(i)}\Pi(j|i)
\;,
\end{equation}
for all $j\in{\mathcal J}$.
It satisfies a kind of modified double stochasticity, namely
\begin{equation}\label{SM8}
\sum_{i\in{\mathcal I}} \pi(j|i)\,d(i)\stackrel{(\ref{SM7})}{=}\sum_{i\in{\mathcal I}}\Pi(j|i)\stackrel{(\ref{SM3b})}{=}\widetilde{d}(j)
\;.
\end{equation}

In accordance with the usual nomenclature of probability theory, functions $X:{\sf E}\rightarrow {\mathbbm R}$ are also called ``random variables".
Their expectation value is defined as
\begin{equation}\label{SM9}
 \langle X\rangle \equiv \sum_{(i,j)\in {\sf E}} X(i,j)\,P(i,j)
 \;,
\end{equation}
if the series converges.
Using a sloppy notation the expectation value will be sometimes also written as $\langle X(i,j)\rangle$ if no misunderstanding is likely to occur.

For the applications we have in mind it is necessary to calculate the expectation value $\langle X(i,j)\rangle$
also if for some points $X(i,j)$ diverges and the probability $P(i,j)$ vanishes. It is not sufficient
to simply exclude these points from the calculation of $\langle X(i,j)\rangle$.
It seems that these mathematical difficulties are connected with the rare events
sampling problem discussed in the literature, see, e.~g., \cite{BS10}.
For our purposes we need only to consider
the points $(i,j)\in{\sf E}$ where $x(i)=0$ and the $ X(i,j)$ are of the form $ X(i,j)=\frac{c(i,j)}{x(i)}$
with some finite numbers $c(i,j)$.
Since  $P(i,j)=\Pi(j|i)\,x(i)$ according to (\ref{SM4a})
the obvious regularisation of the otherwise undefined expectation value will be to cancel $x(i)$ and to
set the contribution of $(i,j)\in{\sf E}$ to the expectation value to $P(i,j)\,X(i,j)=c(i,j)\,\Pi(j|i)$.
Also the above considerations on the conditional probability would have to be reformulated by using this regularisation.
This lends additional meaning to the auxiliary concept of the conditional matrix $\Pi(j|i)$ that has already been
introduced in \cite{SG20} in order to obtain a more symmetric formulation of the framework for sequential measurements.

Taking into account this regularisation procedure we have the following result:
\begin{prop}\label{PropJ}
Under the preceding conditions
the following holds
\begin{equation}\label{SM14}
 \left\langle  \frac{\widetilde{x}(j)}{x(i))}\right\rangle=1
 \;.
\end{equation}
\end{prop}
The proof is elementary, see Appendix \ref{sec:AP}.\\

We will call Eq.~(\ref{SM14}) the ``J-equation" since we think that it contains
the probabilistic core of the Jarzynski equation but should be distinguished from the latter for the sake of clarity.
This claim has been further explained in \cite{SG20}. Due to the symmetry of our assumptions a reciprocal J-equation
could be proven using the second probability distribution $\widetilde{P}(i,j)$, but this will not be needed in what follows.

The probability distributions $q(j)$  and $\widetilde{p}(j)$ defined in (\ref{SM6b}) and (\ref{SM6c}) are completely independent.
A possible specialization of the model for sequential measurements is given by the choice of $\widetilde{x}(j)$ that results in
$\widetilde{p}(j)=q(j)$ for all $j\in{\mathcal J}$, namely
\begin{equation}\label{SM15}
 \widetilde{x}(j) =\frac{q(j)}{\widetilde{d}(j)}
 \;,
\end{equation}
for $\widetilde{d}(j)<\infty$ and  $\widetilde{x}(j) =0$ else.
This will be called the ``minimal case" for reasons to be explained below.

In contrast to \cite{NC00} we will always denote by ``$\log$" the natural logarithm.
Since it is a concave function, Jensen's inequality yields
\begin{equation}\label{SM16}
  \langle \log\,X\rangle \le \log\,\langle X\rangle
\end{equation}
for any random variable $X:{\sf E}\rightarrow{\mathbbm R}_\ge$.
We will define the ``modified Shannon entropy", see \cite{S48}, by
\begin{equation}\label{SM17}
 H(p) \equiv -\sum_i p_i\,\log\frac{p(i)}{d(i)}
 \;,
\end{equation}
and obtain:
\begin{prop}\label{PropH}
\begin{equation}\label{SM18}
  H(p)\le H(q) \le -\sum_{j\in{\mathcal J}}q(j)\,\log\frac{\widetilde{p}(j)}{\widetilde{d}(j)}
  \;.
\end{equation}
\end{prop}
The proof can be found in the Appendix \ref{sec:AP}.
Obviously, $H(q)$ minimises the right hand side of (\ref{SM18}),
thereby justifying the denotation of the choice  (\ref{SM15}) resulting in $\widetilde{p}(j)=q(j)$
as the minimal case.

\section{Applications to quantum theory}\label{sec:G}
\subsection{Sequential measurements in quantum theory}\label{sec:GS}

We consider a quantum system with a Hilbert space ${\mathcal H}$ and a finite number of mutually commuting self-adjoint
operators $\utilde{E}_1,\ldots,\utilde{E}_L$ defined on (suitable domains of) ${\mathcal H}$. They are assumed to have a pure point
spectrum and hence a family of common eigenprojections $(\utilde{P}_i)_{i\in{\mathcal I}}$  such that
\begin{equation}\label{G1}
 \utilde{E}_\lambda=\sum_{i\in{\mathcal I}}E_i^{(\lambda)}\,\utilde{P}_i,\quad \lambda=1,\ldots,L
 \;.
\end{equation}
Here ${\mathcal I}$ is a finite or countable infinite index set to be identified with the outcome set of the first measurement
according to Section \ref{sec:SM}.
In general, the $\utilde{P}_i$ may be of infinite degeneracy; hence we define
\begin{equation}\label{G2}
{\mathcal I}'\equiv \{ i\in{\mathcal I}\left| d(i)\equiv \mbox{Tr}\left( \utilde{P}_i\right)<\infty\right.\}
  \;.
\end{equation}
Further, the $\utilde{P}_i$
are chosen as maximal projections in the sense that $i\neq j$ implies $E_i^{(\lambda)}\neq E_j^{(\lambda)}$ for at least one $\lambda=1,\ldots,L$.
Note the completeness relation
\begin{equation}\label{G3}
 \sum_{i\in{\mathcal I}}\utilde{P}_i={\mathbbm 1}
 \;.
\end{equation}
Physically, the $\utilde{E}_1,\ldots,\utilde{E}_L$ correspond to observables that can be jointly measured. We assume a (mixed) state
of the system before the time $t=t_0$ described by a density operator $\rho_0$ and perform a joint L\"uders measurement, cf.~\cite{BLPY16} (10.22),
of $\utilde{E}_1,\ldots,\utilde{E}_L$
at the time $t=t_0$.
The probability of the outcome $i\in{\mathcal I}$ will be
\begin{equation}\label{G4}
  p(i)=\mbox{Tr} \left( \rho_0\,\utilde{P}_i\right)
  \;,
\end{equation}
satisfying
\begin{equation}\label{G4a}
 \sum_{i\in{\mathcal I}} p(i)=1
  \;.
\end{equation}
In accordance with the remarks after  (\ref{SM6d}) we will make the following
\begin{ass}\label{AS1}
  \begin{equation}\label{AS1a}
    p(i)>0 \mbox{ for all } i\in{\mathcal I}'
    \;.
  \end{equation}
\end{ass}

After the first measurement of the $\utilde{E}_1,\ldots,\utilde{E}_L$ the system is subject to a further time evolution and a
second measurement of (possibly) other observables. Thus the primary preparation together with the first measurement may be considered
as another preparation of a certain state $\rho$, in general different from the initial state $\rho_0$. If a selection according
to a particular outcome $i\in{\mathcal I}'$ is involved this state will be, according to the assumption of a L\"uders measurement,  cf.~\cite{BLPY16} (10.22),
\begin{equation}\label{G5a}
  \rho_i=\frac{\utilde{P}_i\,\rho_0\,\utilde{P}_i}{\mbox{Tr}\left(\rho_0\,\utilde{P}_i \right)}=\frac{\utilde{P}_i\,\rho_0\,\utilde{P}_i}{p(i)}
  \;.
\end{equation}
If no selection according to a particular outcome is involved the state resulting after the first measurement will rather be the
mixed state
\begin{equation}\label{G5b}
  \rho=\sum_{i\in{\mathcal I}'}p(i)\,\rho_i \stackrel{(\ref{G5a})}{=}\sum_{i\in{\mathcal I}'}\utilde{P}_i\,\rho_0\,\utilde{P}_i
  \;.
\end{equation}

In order to apply the results of the preceding section we will make the following crucial assumption
\begin{ass}\label{AS2}
  \begin{equation}\label{AS2a}
    \rho_i=\frac{1}{d(i)}\utilde{P}_i \mbox{ for all } i\in{\mathcal I}'
    \;.
  \end{equation}
\end{ass}
If $\utilde{P}_i$ is a one-dimensional projection, i.~e., if $d(i)=1$, the assumption (\ref{AS2a}) will be automatically
satisfied. In the case of $d(i)>1$ this assumption means that $\rho_0$ is diagonal w.~r.~t.~any common eigenbasis of the
$\utilde{E}_1,\ldots,\utilde{E}_L$. An important case where (\ref{AS2a}) holds is given if $\rho_0$ is a function of the operators
$\utilde{E}_1,\ldots,\utilde{E}_L$, say,
\begin{equation}\label{G5c}
  \rho_0={\mathcal G}\left( \utilde{E}_1,\ldots,\utilde{E}_L\right)
\;.
\end{equation}
For example, the choice of ${\mathcal G}$ as the Boltzmann distribution leads to a Jarzynski equation of the form (\ref{I2}) for $L=1$.

Next we consider a second set of observables described by the mutually commuting self-adjoint operators $\utilde{F}_1,\ldots,\utilde{F}_L$
subject to analogous assumptions. Hence the following holds:
\begin{equation}\label{G6}
 \utilde{F}_\lambda=\sum_{j\in{\mathcal J}}F_j^{(\lambda)}\,\utilde{Q}_j,\quad \lambda=1,\ldots,L
 \;,
\end{equation}
\begin{equation}\label{G8}
 \sum_{j\in{\mathcal J}}\utilde{Q}_j={\mathbbm 1}
 \;,
\end{equation}
\begin{equation}\label{G8a}
 {\mathcal J}'\equiv  \{j\in{\mathcal J}\left| \widetilde{d}(j)\equiv \mbox{Tr}\left( \utilde{Q}_j\right)<\infty\right.\}
   \;,
\end{equation}
and
\begin{ass}\label{AS11}
  \begin{equation}\label{AS11a}
    \widetilde{p}(j)>0 \mbox{ for all } j\in{\mathcal J}'
    \;.
  \end{equation}
\end{ass}
Here $ \widetilde{p}(j)$ denotes an arbitrary probability distribution.

We have chosen another index set ${\mathcal J}$ for the second set of observables in order to stress that no natural identification between both index sets
is required in what follows. Obviously,  ${\mathcal J}$ has to be identified with the second outcome set introduced in Section \ref{sec:SM}.
In general the $\utilde{E}_\lambda$ will not commute with the $\utilde{F}_\mu$. We assume that
a second measurement of the  $\utilde{F}_1,\ldots,\utilde{F}_L$ will be performed at the time $t=t_1>t_0$, not necessarily of L\"uders type.
Between the two measurements in the time interval $(t_0,t_1)$ the evolution of the system can be quite arbitrary and will be described
by a unitary evolution operator $U=U(t_1,t_0)$.\\

In order to apply the results of the last section we will define the quantities $\Pi, x,\widetilde{x}$ and show that
Assumption \ref{axiom} will be satisfied in the quantum case. Moreover, we will show that the probability function $P(i,j)$
has its usual meaning here.

We set
\begin{eqnarray}\label{G9a}
 \Pi(j|i)&=& \mbox{Tr}\left(
 \utilde{Q}_j\,U\,\utilde{P}_i\,U^\ast  \right)\;\mbox{ for all } i\in{\mathcal I},j\in{\mathcal J},\\
  \label{G9b}
  x(i)&=& \frac{p(i)}{d(i)}\;\mbox{ for all } i\in{\mathcal I}' ,\quad \mbox{and}\\
  \label{G9c}
  \widetilde{x}(j)&=& \frac{ \widetilde{p}(j)}{ \widetilde{d}(j)}\;\mbox{ for all } j\in{\mathcal J}'\;.
\end{eqnarray}
It follows that the marginal sums of $\Pi(j|i)$ agree with the degeneracies $d(i)$ and $\widetilde{d}(j)$ defined above.

Moreover,
\begin{eqnarray}\label{G10a}
 P(i,j)&=&\Pi(j|i)\,x(i)\\
 \label{G10b}
 &\stackrel{(\ref{G9a},\ref{G9b})}{=}&
 \mbox{Tr}\left( \utilde{Q}_j\,U\,\utilde{P}_i\,U^\ast \right)\,\frac{p(i)}{d(i)}\\
 \label{G10c}
 &\stackrel{(\ref{AS2a})}{=}&  \mbox{Tr}\left( \utilde{Q}_j\,U\,\rho_i\,U^\ast \right)\,p(i)\\
 \label{G10d}
 &\stackrel{(\ref{G5a})}{=}&  \mbox{Tr}\left( \utilde{Q}_j\,U\,\utilde{P}_i\,\rho_0\,\utilde{P}_i\,U^\ast \right)
\end{eqnarray}
is the correct probability of the outcome $(i,j)$ according to the rules of quantum theory. Moreover,
the following holds:

\begin{prop}\label{LE1}
If the above Assumptions \ref{AS1}, \ref{AS2}, and \ref{AS11} are satisfied then
the quintuple $\left({\mathcal I},{\mathcal J}, \Pi, x, \widetilde{x}\right)$ defined in (\ref{G9a}--\ref{G9c}) also satisfies
 Assumption \ref{axiom} and hence represents a model of sequential measurements.
\end{prop}
The proof can be found in the Appendix \ref{sec:AP}.
Especially, the J-equation (\ref{SM14}) holds in quantum theory as well as the $2^{nd}$ law-like inequality (\ref{SM18}).

\subsection{Results on the von Neumann entropy}\label{sec:GT}

Next we will prove some well-known results connected with the von Neumann entropy using the framework of sequential measurement
sketched in Section \ref{sec:SM}. Recall the definition of the von Neumann entropy
\begin{equation}\label{GT1}
  S(\rho)\equiv -\mbox{Tr }(\rho \log \rho)
  \;,
\end{equation}
for arbitrary statistical operators $\rho$. As usual, the limit $\lim_{x\downarrow 0} x\,\log x=0$ is tacitly understood for
vanishing eigenvalues of $\rho$.

For this subsection we will slightly specialise the definitions of the preceding subsection \ref{sec:GS}.
We note that the eigenvalues of the operators $\utilde{E}_\lambda$ and $\utilde{F}_\lambda$ to be measured
do not enter into the scheme of sequential measurement but only the corresponding eigenprojections.
We use this freedom of choosing the eigenvalues in the following way.
Let $\rho$ and $\sigma$ be two statistical operators
with respective spectral decompositions and traces
\begin{eqnarray}
\label{PK1a}
  \rho &=& \sum_{i\in{\mathcal I}}r_i\,\utilde{P}_i\;, \\
  \label{PK1b}
  1 &=&\mbox{Tr }\rho=  \sum_{i\in{\mathcal I}}r_i\,\mbox{Tr}\utilde{P}_i
  \equiv  \sum_{i\in{\mathcal I}}r_i\,d(i) \equiv  \sum_{i\in{\mathcal I}}p(i)\;, \\
  \label{PK1c}
   \sigma &=& \sum_{j\in{\mathcal J}}s_j\,\utilde{Q}_j\;, \\
   \nonumber
  1 &=&\mbox{Tr }\sigma=  \sum_{j\in{\mathcal J}}s_j\,\mbox{Tr}\utilde{Q}_j
  \equiv  \sum_{j\in{\mathcal J}}s_j\,\widetilde{d}(j) \equiv  \sum_{j\in{\mathcal J}}\widetilde{p}(j)\;. \\
  && \label{PK1d}
\end{eqnarray}

Since $\rho$ and $\sigma$ are Hermitean operators they can also be viewed as observables.
Thus we choose the initial state $\rho_0=\rho$ and perform a first L\"uders measurement of $\rho$
at time $t=t_0$ with outcome $i\in{\mathcal I}$. The state after this measurement without selection
is obviously again $\rho$. It follows that the condition
(\ref{AS2a}) will be automatically satisfied.
Between $t=t_0$ and $t=t_1>t_0$ no interaction takes place, i.~e., $U(t_1,t_0)={\mathbbm 1}$.
Then at time $t=t_1$ a second measurement of $\sigma$ is performed with outcome $j\in{\mathcal J}$.
The assumption $U(t_1,t_0)={\mathbbm 1}$  does not imply any loss of generality since $\sigma$ is completely arbitrary.
The conditional matrix (\ref{G9a}) of the sequential measurement will assume the simplified form
\begin{equation}\label{GT2}
 \Pi(j|i)=\mbox{Tr}\left( \utilde{P}_i\, \utilde{Q}_j \right)
\;.
\end{equation}
Moreover, the abstract eigenvalues can be identified with the actual eigenvalues of $\rho$ and $\sigma$, i.~e.,
\begin{equation}\label{GT3}
 x(i)=r_i\; \mbox{ and } \widetilde{x}(j)=s_j\;\mbox{ for all } i\in{\mathcal I} \mbox{ and }j\in{\mathcal J}
 \;.
\end{equation}

For arbitrary statistical operators $\rho,\sigma$ the ``relative entropy" is defined as
\begin{equation}\label{GT4}
  S(\rho || \sigma)\equiv \mbox{Tr }(\rho \log \rho) -\mbox{Tr }(\rho \log \sigma)
  \;,
\end{equation}
compare \cite{NC00}, (11.50). The relative entropy may diverge for certain choices of $\rho$ and $\sigma$, see \cite{L73}.
It is never negative according to
\begin{prop}\label{PropKlein}
  {\bf (Klein's inequality)}\\
  \begin{equation}\label{GT5}
  S(\rho || \sigma)\ge 0
  \;,
  \end{equation}
\end{prop}
compare \cite{K31} and, for a more recent reference, \cite{NC00}, Theorem 11.7.
Our alternative proof using sequential measurements can be found in Appendix \ref{sec:AP}.\\

For the remainder of this subsection we will concentrate on the case where the statistical operator $\sigma$ is chosen
in such a way that the ``minimal case" according to (\ref{SM15}) is obtained. More precisely, we define
\begin{defi}\label{D1}
 The pair $(\rho,\sigma)$ of statistical operators will be called ``minimal" iff
 \begin{equation}\label{D1a}
  \widetilde{p}(j)=\mbox{Tr}\left( \sigma\,\utilde{Q}_j\right)=\mbox{Tr}\left( \rho\,\utilde{Q}_j\right)=q(j),
 \end{equation}
 for all $j\in{\mathcal J}$, where the $\utilde{Q}_j$ are the eigenprojections of $\sigma$ according to (\ref{PK1c}).
\end{defi}
In this case Eq.~(\ref{SM18}) reduces to $H(p)\le H(q)$ and the right hand side of this inequality can be identified with
$S(\sigma)$:
\begin{prop}\label{Prop2nd}
 If $(\rho,\sigma)$ is minimal then $S(\rho || \sigma)=S(\sigma)-S(\rho)$ and hence, by Proposition \ref{PropKlein},  $S(\rho)\le S(\sigma)$.
\end{prop}
The proof can be found in Appendix \ref{sec:AP}.\\

It is worthwhile noting that the converse of this Proposition does not hold. We will present a counter-example
where $S(\rho || \sigma)=S(\sigma)-S(\rho)$ without $(\rho,\sigma)$ being minimal.
The Hilbert space of this counter-example will be ${\mathcal H}={\mathbbm C}^4\cong {\mathbbm C}^2\otimes {\mathbbm C}^2$ and
$\rho$ will be the projector onto an entangled state $\phi$:
\begin{eqnarray}
\label{GT6a}
  \rho &=& P_\phi,\quad \mbox{where }\phi=\frac{1}{2}\left( \uparrow\downarrow+\sqrt{3}\downarrow\uparrow\right)\;, \\
  \sigma &=& \rho_1\otimes \rho_2\;,
\end{eqnarray}
and the $\rho_i,\,i=1,2$ denoting the partial traces of $\rho$ w.~r.~t.~the tensor factors of ${\mathbbm C}^4\cong {\mathbbm C}^2\otimes {\mathbbm C}^2$.
Note that the $\rho_i,\,i=1,2$ are isospectral since $\rho$ is a pure state admitting a Schmidt decomposition.
From this it follows that two eigenvalues of $\sigma$ will be degenerate.
Moreover, for the above choice of $\sigma$ the equation $S(\rho || \sigma)=S(\sigma)-S(\rho)$ holds in general, see the proof of the
subadditivity of the von Neumann entropy in \cite{NC00}, 11.3.4.  In our case $\sigma$ is diagonal in the standard basis of
${\mathbbm C}^4$:
\begin{equation}\label{GT7}
  \sigma=\left(
\begin{array}{cccc}
 \frac{3}{16} & 0 & 0 & 0 \\
 0 & \frac{1}{16} & 0 & 0 \\
 0 & 0 & \frac{9}{16} & 0 \\
 0 & 0 & 0 & \frac{3}{16} \\
\end{array}
\right)
\;,
\end{equation}
whereas
\begin{equation}\label{GT8}
  \rho=\left(
\begin{array}{cccc}
 0 & 0 & 0 & 0 \\
 0 & \frac{1}{4} & \frac{\sqrt{3}}{4} & 0 \\
 0 & \frac{\sqrt{3}}{4} & \frac{3}{4} & 0 \\
 0 & 0 & 0 & 0 \\
\end{array}
\right)
\;.
\end{equation}
Obviously, the $q(j)$, the diagonal entries of $\rho$, are different from the $\widetilde{p}(j)$, the
diagonal entries of $\sigma$, and hence the pair $(\rho,\sigma)$ is not minimal.\\

Next we turn to the problem how the von Neumann entropy of a state changes during a quantum measurement.
Obviously, this problem depends on the theoretical description of state changes during measurements and hence
leads to the notions of operations and instruments, see  \cite{BLPY16}.
The simplest case is that of a L\"uders measurement $\mathfrak{I}$, see \cite{BLPY16}, (10.22),
\begin{equation}\label{GT9}
 \rho\mapsto  \mathfrak{I}(n)(\rho)=\utilde{P}_n\,\rho\,\utilde{P}_n
  \;,
\end{equation}
where $\left(\utilde{P}_n\right)_{n\in{\mathbbm N}}$ is a complete system of orthogonal projections,
not necessarily finite-dimensional ones.
The state change without any selection will be the trace preserving map
\begin{equation}\label{GT10}
 \rho\mapsto \sigma\equiv \sum_{n\in{\mathbbm N}}\utilde{P}_n\,\rho\,\utilde{P}_n
  \;.
\end{equation}
It is well-known, see \cite{vN32} and \cite{NC00}, theorem 11.9, that L\"uders measurements increase entropy:
\begin{prop}\label{PropLMIE}
With the preceding definitions the following holds:
\begin{equation}\label{LMIE1}
 S(\rho)\le S(\sigma)
 \;.
\end{equation}
\end{prop}
A proof using the framework of sequential measurement can be found in Appendix \ref{sec:AP}.
We stress that the hypothetical sequential measurement used in this proof and the original L\"uders measurement (\ref{GT9})
are different, although related. This will be underscored by the following remarks:
\begin{itemize}
 \item The first measurement of $\rho$ used in the proof with outcome $i\in{\mathcal I}$ is not part of the L\"uders measurement.
 \item The projections $\utilde{Q}_j$ of the second hypothetical measurement used in the proof are finite-dimensional in contrast to the $\utilde{P}_n$,
 they rather represent a refinement of the family $\left(\utilde{P}_n\right)_{n\in{\mathbbm N}}$.
 \item  Moreover, the second hypothetical measurement used in the proof need not be of L\"uders type.
\end{itemize}

In the case of a  more general instrument than that of L\"uders type it is well-known that a statement analogous to
(\ref{LMIE1}) may fail, i.~e., a generalised measurement can decrease entropy, see \cite{NC00}, exercise 11.15.
This may sound paradoxical at first sight but can be understood by considering the ``measurement dilation" of a general
instrument, see \cite{BLPY16}, chapter 7.7. This means that the object system with Hilbert space ${\mathcal H}_1$ is coupled
to a second system (``measuring device") with Hilbert space ${\mathcal H}_2$ and, after some interaction of the total system
described by a unitary evolution operator $U$, a L\"uders measurement with projectors $\utilde{P}_n$ is performed at the measuring device.
The final step of the state change consists of a partial trace $\mbox{Tr}_2$ that yields a mixed state $\sigma$ of the object system.
We thus obtain for the state change without selection the following expression
\begin{equation}\label{GT11}
  \rho\mapsto\sigma = \mbox{Tr}_2
  \sum_{n\in{\mathbbm N}}
  \left(   {\mathbbm 1}\otimes \utilde{P}_n\right) \, U\,\left(\rho\otimes P_\phi\right)\,U^\ast \, \left( {\mathbbm 1}\otimes \utilde{P}_n \right)
  \;,
\end{equation}
where we have used that the initial state of the measuring device can be chosen as a pure state $P_\phi$, see \cite{BLPY16}, chapter 7.7.
Without the partial trace $\mbox{Tr}_2$ this can be understood as a L\"uders measurement of the total system with projectors
\begin{equation}\label{GT12}
 \utilde{Q}_n\equiv U^\ast \left( {\mathbbm 1}\otimes \utilde{P}_n \right) \,U
\end{equation}
and the initial state $\rho\otimes P_\phi$. It follows from Proposition \ref{PropLMIE} that the total entropy does not decrease, i.~e.,
\begin{equation}\label{GT13}
S_1\equiv S(\rho)=S(\rho\otimes P_\phi)\le S\left(\rho' \right)\equiv S_2
 \;,
\end{equation}
where
\begin{equation}\label{GT14}
 \rho' \equiv \sum_{n\in{\mathbbm N}}  \utilde{Q}_n\,\left(\rho\otimes P_\phi \right)\, \utilde{Q}_n
  \;,
\end{equation}
and we have used the fact that the entropy of a tensor product is additive and the entropy of a pure state vanishes.
By forming partial traces the total entropy further increases, see \cite{NC00}, chapter 11.3.4, and thus
\begin{equation}\label{GT15}
 S_1\le S_2\le S\left( \mbox{Tr}_1 (\rho')\right)+S\left( \mbox{Tr}_2 (\rho')\right)\equiv S_{32}+S_{31}\equiv S_3
 \;.
\end{equation}

Hence the total entropy during a generalised measurement does not decrease if the possible entropy increase of the measuring device
is taken into account. The mentioned counter examples of a decreasing entropy of the object system occur if
$S_{31}<S_1$ which is well possible in spite of $S_1\le S_{31}+S_{32}$. This means that the decrease of the entropy of the object system
must be (over)compensated by an increase of the measurement device's entropy,
see the corresponding discussion in \cite{L73} and \cite{BLM96}, chapter III.5.

\section{Summary and Outlook}
\label{sec:SO}
A general framework for sequential measurement including the J-equation has been recently formulated and shown to
be realised in quantum theory \cite{SG20}. The J-equation comprises the various variants of the famous Jarynski equation.
In the present paper we have slightly generalised this framework
in order to cope with the problem of vanishing probabilities. A standard application of the J-equation results from
Jensen's inequality and the fact that the logarithm is a concave function. The resulting inequality has been shown to be essentially
equivalent to Klein's inequality already derived in $1931$ in the context of quantum thermodynamics.
This opens an unexpected connection between non-equilibrium quantum statistical mechanics and general entropy theory
mainly used in context with quantum information, see \cite{NC00}. The concept of sequential measurements is proving fruitful
not only in its direct application to sequential measurements but also in the sense of a mathematical tool.
Thus the new proofs of well-known laws like Proposition \ref{PropLMIE} opens up a new perspective in this field,
insofar as the various $2^{nd}$ law-like statements can be viewed as consequences of an underlying J-equation.
It would be desirable to use these tools to simplify the involved proofs of, say, the strong subadditivity of the von Neumann entropy,
see \cite{NC00}, chapter $11.4$, but this is definitively outside the scope of the present paper.

\appendix

\section{Proofs}
\label{sec:AP}

\noindent {\bf Proof} of Proposition \ref{PropJ}:
\begin{eqnarray}\label{SM15a}
  \left\langle  \frac{\widetilde{x}(j)}{x(i))}\right\rangle
  &\stackrel{(\ref{SM9})}{=}&
  \sum_{(i,j)\in{\sf E}}P(i,j)\,\frac{\widetilde{x}(j)}{x(i))}\\
  \label{SM15b}
  &\stackrel{(\ref{SM4a})}{=}&
  \sum_{(i,j)\in{\sf E}}\Pi(j|i)\,\widetilde{x}(j)\\
   \label{SM15c}
  &\stackrel{(\ref{SM4b})}{=}&
    \sum_{(i,j)\in{\sf E}}\widetilde{P}(j,i)
        \stackrel{(\ref{SM5b})}{=}1
     \;.
\end{eqnarray}
We would like to point out that the summations in (\ref{SM15a}) - (\ref{SM15c}) have to be extended over the whole domain
${\sf E}$ including those points where $x(i)$ vanishes. The contribution to the expectation value from these points has
to be calculated according to the regularisation procedure explained  in the text following (\ref{SM9}).
\hfill$\Box$\\

\noindent {\bf Proof} of Proposition \ref{PropH}:
\begin{eqnarray}
\label{PH1a}
  0 &=& \log\,1\stackrel{(\ref{SM14})}{=}\log \left\langle  \frac{\widetilde{x}(j)}{x(i)}\right\rangle\\
  \label{PH1b}
   &\stackrel{(\ref{SM15})}{\ge}&\left\langle \log\left(  \frac{\widetilde{x}(j)}{x(i)}\right)\right\rangle\\
   \label{PH1c}
   &=& \left\langle \log \widetilde{x}(j) \right\rangle-\left\langle \log x(i) \right\rangle\\
   \nonumber
     &=& \sum_{(i,j)\in{\sf E}}P(i,j)\,\log\widetilde{x}(j) -\sum_{(i,j)\in{\sf E}}P(i,j)\,\log x(i)\\
      \label{PH1d} &&\\
      \label{PH1e}
   &\stackrel{(\ref{SM6a},\ref{SM6b})}{=}& \sum_{j\in{\mathcal J}}q(j)\,\log\widetilde{x}(j) -\sum_{i\in{\mathcal I}}p(i)\,\log x(i)\\
   \label{PH1f}
   &\stackrel{(\ref{SM6c},\ref{SM17})}{=}&   \sum_{j\in{\mathcal J}}q(j)\,\log\frac{\widetilde{p}(j)}{\widetilde{d}(j)} +H(p)
    \;.
\end{eqnarray}
Hence
\begin{equation}\label{PH2}
H(p)\le -  \sum_{j\in{\mathcal J}}q(j)\,\log\frac{\widetilde{p}(j)}{\widetilde{d}(j)}
\;,
\end{equation}
and especially for the minimal case
\begin{equation}\label{PH3}
H(p)\le H(q)
\;.
\end{equation}
It remains to show the second inequality in (\ref{SM18}).
The $\log$ function is bounded by its tangent at $x=1$:
\begin{equation}\label{PH4}
  \log x \le x-1\quad \mbox{for } x>0
  \;,
\end{equation}
and thus
\begin{eqnarray}
 \label{PH5a}
  &&-\sum_{j\in{\mathcal J}}q(j)  \log \left(\frac{\widetilde{p}(j)}{\widetilde{d}(j)}\frac{\widetilde{d}(j)}{q(j)}\right)\\
   \label{PH5b}
 &\ge& -\sum_{j\in{\mathcal J}}q(j)
  \left(\frac{\widetilde{p}(j)}{\widetilde{d}(j)}\frac{\widetilde{d}(j)}{q(j)}-1\right)\\
   \label{PH5c}
    &=&  -\sum_{j\in{\mathcal J}}\widetilde{p}(j)+\sum_{j\in{\mathcal J}}q(j)=-1+1=0
  \;.
\end{eqnarray}
This entails
\begin{equation}\label{PH6}
 H(q)= -\sum_{j\in{\mathcal J}}q(j)\log \frac{q(j)}{\widetilde{d}(j)}\le  -\sum_{j\in{\mathcal J}}q(j)\log \frac{\widetilde{p}(j)}{\widetilde{d}(j)}
 \;,
\end{equation}
thereby completing the proof of Proposition \ref{PropH}. \hfill$\Box$\\

\noindent {\bf Proof} of Proposition \ref{LE1}:
We have to prove (\ref{SM5a}) and (\ref{SM5b}).
To this end we consider
\begin{eqnarray}
\nonumber
 \sum_{(i,j)\in{\sf E}}P(i,j)&=&\sum_{i\in{\mathcal I}'} \mbox{Tr}\left(
 \left( \sum_{{j\in{\mathcal J}}}
  \utilde{Q}_j
 \right)
  \,U\,\utilde{P}_i\,U^\ast
 \right)\,\frac{p(i)}{d(i)} \\
 \label{G11a}
   &\stackrel{(\ref{G8})}{=}& \sum_{i\in{\mathcal I}'} \mbox{Tr}\left(
   \,U\,\utilde{P}_i\,U^\ast
 \right)\,\frac{p(i)}{d(i)}\\
 \nonumber
 &=&\sum_{i\in{\mathcal I}'}\frac{ \mbox{Tr}\left(
 \utilde{P}_i
 \right)}{d(i)}\,p(i)\stackrel{(\ref{G2})}{=}\sum_{i\in{\mathcal I}'}p(i)=1
 \;.\\
 \label{G11b}
\end{eqnarray}

Further,
\begin{eqnarray}
\label{G12a}
   \sum_{(i,j)\in{\sf E}}\widetilde{P}(i,j)&=& \sum_{(i,j)\in{\sf E}}\Pi(i,j)\,\widetilde{x}(j)\\
   \nonumber
   &\stackrel{(\ref{G9c})}{=}&\sum_{j\in{\mathcal J}'} \mbox{Tr}\left(
  \utilde{Q}_j
  \,U\,
  \left( \sum_{i\in{\mathcal I}}
  \utilde{P}_i
  \right)
  \,U^\ast
 \right)\,\frac{\widetilde{p}(j)}{\widetilde{d}(j)} \\
 \nonumber
 &\stackrel{(\ref{G3})}{=}&\sum_{j\in{\mathcal J}'}\frac{ \mbox{Tr}\left(
 \utilde{Q}_j
 \right)}{\widetilde{d}(j)}\,\widetilde{p}(j)\stackrel{(\ref{G8a})}{=}\sum_{j\in{\mathcal J}'}\widetilde{p}(j)=1
 \;,\\
 \label{G12b}
\end{eqnarray}
which completes the proof of Proposition \ref{LE1}. \hfill$\Box$\\

\noindent {\bf Proof} of Proposition \ref{PropKlein}:
With the definitions of Section \ref{sec:GT} we obtain
\begin{eqnarray}
\nonumber
  S(\rho) &=& -\mbox{Tr}\left( \rho\,\log\rho\right)\stackrel{(\ref{PK1a})}{=}-\sum_{i\in{\mathcal I}'} r_i \log(r_i)\,\mbox{Tr } \utilde{P}_i\\
  \label{PK3a}&&\\
  \label{PK3b}
   &\stackrel{(\ref{PK1b})}{=}& -\sum_{i\in{\mathcal I}'}p(i) \log\frac{p(i)}{d(i)}=H(p)
   \;,
\end{eqnarray}
and further
\begin{eqnarray}
\label{PK4a}
 \mbox{Tr}\left( \rho\,\log\sigma\right) &\stackrel{(\ref{PK1a},\ref{PK1c})}{=}&\sum_{(i,j)\in{\sf E}} r_i \log(s_j)\,\mbox{Tr}\left( \utilde{P}_i\,\utilde{Q}_j\right)\\
 \label{PK4b}
   &\stackrel{(\ref{GT2},\ref{GT3})}{=}&\sum_{(i,j)\in{\sf E}} x(i) \log(\widetilde{x}(j))\,\Pi(j|i)\\
      \label{PK4c}
   &\stackrel{(\ref{SM6b})}{=}&\sum_{j\in {\mathcal J}}q(j)\,\log(\widetilde{x}(j))  \\
    \label{PK4d}
   &\stackrel{(\ref{SM6c})}{=}&\sum_{j\in {\mathcal J}}q(j)\,\log\left(\frac{\widetilde{p}(j)}{\widetilde{d}(j)}  \right)
   \;.
\end{eqnarray}
This implies
\begin{eqnarray}\label{PK5a}
  S(\rho||\sigma) &=& -S(\rho)-\mbox{Tr}(\rho\log\sigma) \\
  \nonumber
   &\stackrel{(\ref{PK3b},\ref{PK4d})}{=}& -H(p)-\sum_{j\in {\mathcal J}}q(j)\,\log\left(\frac{\widetilde{p}(j)}{\widetilde{d}(j)}  \right)
   \stackrel{(\ref{SM18})}{\ge} 0\;,\\
   \label{PK5b} &&
   \end{eqnarray}
and the proof of Proposition \ref{PropKlein} is complete. \hfill$\Box$\\

\noindent {\bf Proof} of Proposition \ref{Prop2nd}:
It suffices to show that $-S(\sigma)= \mbox{Tr} \left( \rho\,\log \sigma\right)$:
\begin{eqnarray}
\label{PP1a}
 -S(\sigma) &\stackrel{(\ref{PK1c})}{=}& \sum_{j\in{\mathcal J}}s_j\,\log\,s_j\; \mbox{Tr}\, \utilde{Q}_j \\
   &\stackrel{(\ref{PK1d})}{=}& \sum_{j\in{\mathcal J}}\frac{\widetilde{p}(j)}{\widetilde{d}(j)}\,\log\,s_j\; \widetilde{d}(j)\\
   &=& \sum_{j\in{\mathcal J}}\widetilde{p}(j) \,\log\,s_j\\
   &\stackrel{(\ref{D1a})}{=}&\sum_{j\in{\mathcal J}}q(j) \,\log\,s_j\\
   &\stackrel{(\ref{SM6b})}{=}&\sum_{j\in{\mathcal J}}\left( \sum_{i\in{\mathcal I}}P(i,j)\right) \,\log\,s_j\\
   &\stackrel{(\ref{SM4a})}{=}&\sum_{j\in{\mathcal J}}\left( \sum_{i\in{\mathcal I}}\Pi(j|i)x(i)\right) \,\log\,s_j\\
   &\stackrel{(\ref{GT2}),(\ref{GT3})}{=}&\sum_{(i,j)\in{\sf E}}r_i\,\log\,s_j\,\mbox{Tr} \left( \utilde{P}_i  \utilde{Q}_j\right)\\
   &\stackrel{(\ref{PK1a}),(\ref{PK1c})}{=}& \mbox{Tr}\left(\rho\,\log\,\sigma \right)\;.
\end{eqnarray}
This completes the proof of Proposition \ref{Prop2nd}. \hfill$\Box$\\

\noindent {\bf Proof} of Proposition \ref{PropLMIE}:
The claim (\ref{LMIE1}) follows from Proposition \ref{Prop2nd} if we can show that the pair $(\rho,\sigma)$ is minimal in the sense of
Definition \ref{D1}. If
\begin{equation}\label{PL1}
 \sigma=\sum_{n\in{\mathbbm N}}\utilde{P}_n\,\rho \utilde{P}_n=\sum_{j\in{\mathcal J}}s_j\,\utilde{Q}_j
\end{equation}
is the spectral decomposition of $\sigma$ it follows that all $\utilde{Q}_j$ commute with all $\utilde{P}_n$
for $j\in{\mathcal J}$ and $n\in{\mathbbm N}$. It follows that
\begin{eqnarray}
\label{PL2a}
\widetilde{p}(j)&=& \mbox{Tr}\left( \sigma\,\utilde{Q}_j\right) \\
\label{PL2b}
   &=&  \sum_{n\in{\mathbbm N}}\mbox{Tr}\left(\utilde{P}_n\,\rho \utilde{P}_n\,\utilde{Q}_j\right)\\
   \label{PL2c}
   &=&  \sum_{n\in{\mathbbm N}}\mbox{Tr}\left(\utilde{P}_n\,\rho \,\utilde{Q}_j\, \utilde{P}_n\right)\\
   \label{PL2d}
   &=& \mbox{Tr}\left( \rho\,\utilde{Q}_j\left( \sum_{n\in{\mathbbm N}}\utilde{P}_n \right)\right)\\
   \label{PL2e}
   &=& \mbox{Tr}\left( \rho\,\utilde{Q}_j\right)=q(j)
   \;,
\end{eqnarray}
where we have used  $[\utilde{Q}_j,\utilde{P}_n]=0$ in (\ref{PL2c}) and $\sum_{n\in{\mathbbm N}}\utilde{P}_n ={\mathbbm 1}$ in (\ref{PL2e}).
Hence the Definition \ref{D1} of a minimal pair $(\rho,\sigma)$ is satisfied
and the proof of Proposition \ref{PropLMIE} is complete. \hfill$\Box$\\

\begin{acknowledgments}
This work has been funded by the Deutsche
Forschungsgemeinschaft (DFG), Grant No. 397107022
(GE 1657/3-1) within the DFG Research Unit FOR 2692
We thank all members of this research unit for stimulating and
insightful discussions and hints to relevant literature.
One of us (H.-J.S.) wants to gratefully acknowledge the fruitful discussions with
Paul Busch during the last years before his deplorable and untimely passing away.
\end{acknowledgments}


\end{document}